# EL INSTITUTO DE FÍSICA DE LA PLATA EN LA PRIMERA POSGUERRA: LA TRANSMISIÓN CALIFICADA DE LAS CIENCIAS A LA JUVENTUD

Alejandro Gangui
Universidad de Buenos Aires
CONICET - Universidad de Buenos Aires, Instituto de Astronomía y Física del Espacio

Eduardo L. Ortiz
Imperial College London, UK.

**RESUMEN:** En la primera década del siglo XX la Universidad Nacional de La Plata, fundada en 1905, creó un moderno y bien dotado Instituto de Física. En este trabajo estudiamos el impacto que esa iniciativa tuvo sobre la modernización de la enseñanza de la física a nivel medio y universitario en la Argentina. Nos concentramos en dos de los egresados más representativos del Instituto de aquellos años, Ramón G. Loyarte y Enrique Loedel Palumbo, y analizamos sus trabajos pedagógicos más importantes y la recepción pública que tuvieron. Estos trabajos son una muestra del aporte del Instituto de Física a la elevación del nivel de educación nacional en el campo de las ciencias físicas en la primera mitad del siglo XX.

**PALABRAS CLAVE**: Ramón Loyarte; Enrique Loedel Palumbo; enseñanza de la física; textos de física; primera mitad del siglo XX.

## THE INSTITUTE OF PHYSICS OF LA PLATA IN THE FIRST POST-WAR PERIOD: THE QUALIFIED TRANSMISSION OF SCIENCES TO YOUNG PEOPLE

**ABSTRACT:** In the first decade of the 20th century, the Universidad Nacional de La Plata, founded in 1905, created a modern and well-equipped Physics Institute. In this paper we study the impact that this initiative had on the modernisation of physics teaching at secondary and university level in Argentina. We focus on two of the most representative graduates of the Institute in those years, Ramón G. Loyarte and Enrique Loedel Palumbo, and analyse their most important pedagogical works and the public reception they received. These works are a sample of the contribution of the Institute of Physics to the raising of the level of national education in the field of physical sciences in the first half of the 20th century.

**KEY WORDS:** Ramón Loyarte; Enrique Loedel Palumbo; physics teaching; physics texts; first half of the 20th century.

# INTRODUCCIÓN

La apreciación, quizás algo difusa, de la importancia que podría tener para la Argentina el desarrollo de la física moderna está presente desde los primeros años del siglo XX. Sin duda, la telegrafía sin hilos, la radioactividad, los rayos X y otros asombrosos inventos y descubrimientos contribuyeron a alimentar ese interés. La fundación del Instituto de Física de la Universidad Nacional de La Plata (UNLP) y la incorporación de destacados especialistas extranjeros para conducirlo refuerzan esa hipótesis (ver el capítulo de reseña de von Reichenbach y Bibiloni, 2012).

La creación del Instituto en la moderna ciudad de La Plata fue ampliamente comentada por la prensa local y nacional, generalmente en forma muy positiva; su desarrollo fue seguido con el mismo interés. Muy tempranamente alcanzó, incluso, a la primera plana de periódicos de circulación amplia. En particular, estos medios informaron a sus lectores sobre las conferencias públicas dictadas en ese instituto, ilustradas con asombrosos experimentos de física, y dieron noticias del arribo al país de nuevos físicos, contratados en Europa.

Sin embargo, este no fue un hecho aislado. A principios de la década de 1920 Jorge Duclout (1853-1929), un destacado científico local formado en el ETH de Zurich, propuso al Consejo Superior de la Universidad de Buenos Aires que se invitara a Albert Einstein (1879-1955) a visitar la Argentina (Ortiz, 1995; Gangui y Ortiz, 2014). Su proyecto tuvo una recepción entusiasta. Sin duda se trataba de una empresa compleja que, para la Universidad de Buenos Aires, suponía una inversión económica y logística no despreciable. Difícilmente hubiera sido posible esa invitación sin una percepción, quizás imprecisa pero firme, de la importancia que la incorporación de los resultados de la nueva física podría significar para la Argentina.

Años más tarde, algunos de los egresados del Instituto de Física hicieron contribuciones científicas de valor a esa disciplina, dentro de las limitaciones de la Argentina de su época. De hecho, aquel grupo de científicos definió en su país, particularmente en la primera mitad del siglo XX, el concepto de investigación en física moderna (Pyenson, 1985; von Reichenbach y Bibiloni, 2012).

En este trabajo nos interesa profundizar en ciertos aspectos que van más allá del avance de la ciencia universitaria: argumentamos que, independientemente de aquellas realizaciones en el campo de la investigación, los primeros graduados de ese instituto fueron también protagonistas principales en el proceso de comunicar las ideas de su ciencia a un público más amplio que el de los especialistas. En particular, fueron los responsables de promover la modernización de la enseñanza de la física en las escuelas medias y universitarias.

Efectivamente, a partir de mediados de la década de 1920 las contribuciones pedagógicas de egresados del Instituto de Física cubren un abanico amplio que va desde las obras de carácter elemental hasta las que atienden a las necesidades de los estudiantes universitarios que cursan estudios en ramas de las ciencias exactas y de la ingeniería.

Además, algunos de los físicos formados en La Plata expresaron un interés activo por la historia y por la filosofía de la física, y se preocuparon también por contribuir al estudio histórico-crítico del proceso de transmisión de su ciencia a nivel de la enseñanza media.

Argumentamos en este trabajo que, a lo largo de la primera mitad del siglo XX, dentro del grupo amplio de esfuerzos realizados para la creación y mantenimiento del Instituto de Física de la UNLP, que supusieron una contribución económica substancial por parte del Estado Nacional, el aporte a la elevación del nivel de educación nacional en el campo de las ciencias físicas se destaca como uno de los logros más claramente tangibles. Finalmente, tratamos de destacar que esas actividades se inscriben dentro del esfuerzo modernizador de la cultura argentina promovido por el Dr. Joaquín V. González (1863-1923), fundador y presidente de la UNLP, durante sus primeros cuatro períodos, entre 1906 y 1918 (Nazar Anchorena, 1927, 63, 75).

## LA PRIMERA POSGUERRA Y LA MODERNIZACIÓN DE LA ENSEÑANZA DE LAS CIENCIAS A ESCALA NACIONAL: ESFUERZOS REALIZADOS DURANTE LA PRESIDENCIA DE ALVEAR

En esta sección nos ocuparemos de los esfuerzos realizados para renovar y actualizar los tradicionales *Apuntes de Física* de Tebaldo J. Ricaldoni (1873-1923) (Ricaldoni, 1896), que fue uno de los textos de física más ampliamente usados en la Argentina entre fines del siglo XIX y fines de la década de 1910 (von Reichenbach y Cappannini, 2016). Esa obra fue repetidamente actualizada, para adecuarla a las sucesivas reformas introducidas en los planes de estudio de los establecimientos de enseñanza media.

Efectivamente, durante las dos primeras décadas del siglo XX la configuración del sistema educativo fue revisada por los sucesivos ministros de Instrucción Pública. Los intentos más significativos de reforma se deben a los esfuerzos de Antonio Bermejo (1853-1929) en 1897, Osvaldo Magnasco (1864-1920) en 1899, a la encuesta informativa promovida por Rómulo S. Naón (1875-1941) en 1909, al plan de Juan M. Garro (1847-1927) en 1912 y al plan

de Ernesto Nelson (1873-1959) de 1915 (ver Saavedra Lamas, 1916, y también Puiggrós, 2002, y Solari, 2002).

En esos planes se discutió la posición relativa de diferentes formas de educación secundaria: humanista, normal, técnica, comercial, industrial, militar, etc., y se intentó, con diverso éxito, revisar la posición de las ciencias en la enseñanza secundaria.

Las reformas, e intentos de reforma, continuaron en la década siguiente; entre 1922 y 1928, durante la Presidencia del Dr. Marcelo T. de Alvear (1868-1942), su ministro de Justicia e Instrucción Pública, el Dr. Antonio Sagarna (1874-1949) impulsó una nueva reforma de los planes de estudio, tanto a nivel primario como secundario. Su objetivo era lograr una enseñanza más estrechamente basada en métodos y conceptos científicos modernos (Zusman, 1996, 184, nota 21; sobre la discusión parlamentaria ver Cámara de Diputados, 1926).

En 1924 Ramón G. Loyarte (1888-1944) fue invitado por el Gobierno Nacional para actuar como asesor *ad-honorem* del Ministerio de Justicia e Instrucción Pública. Loyarte era uno de los ex-alumnos más destacados del profesor Emil Bose (1874-1911), el primer físico alemán contratado para dirigir el Instituto de La Plata, y tenía un interés firme en la modernización de la enseñanza de las ciencias, y en particular de la física. Dos años más tarde, el ministro Sagarna lo designó miembro de una Comisión Ministerial que él acababa de crear para estudiar, a nivel nacional, la posible renovación y modernización de los textos de enseñanza secundaria. Nuevamente, la función de Loyarte era la de asesorar en el área de la física. En ese momento Loyarte había recibido ya el Premio Nacional de Ciencias, que en esa oportunidad se otorgó por primera vez a un especialista en las ciencias físicas. En 1925, luego del retiro del segundo director alemán Richard Gans (1880-1954), Loyarte había sido designado director del Instituto de Física.

Su pertenencia a aquella Comisión Ministerial le permitió a Loyarte contribuir a decidir acerca del enfoque que debería ofrecer un nuevo texto de física a nivel de la enseñanza media y sobre los temas que éste debía cubrir.

Sin duda, Loyarte estaba en excelentes condiciones para materializar esas ideas a través de un nuevo texto de física elemental, adaptado a aquellos nuevos diseños curriculares. Sin embargo, no abordó la tarea solo; como colaborador para la redacción de esa obra invitó a su joven colega Enrique Loedel Palumbo (1901-1962), un físico de origen uruguayo que había sido discípulo de Gans en La Plata.

Loedel Palumbo reunía condiciones muy particulares, que la reciente visita de Albert Einstein a la Argentina (Ortiz, 1995; Gangui y Ortiz, 2008; 2020) había puesto de manifiesto. En un coloquio con Einstein, auspiciado por la Academia de Ciencias en Buenos Aires, Loedel

Palumbo se mostró como un interlocutor articulado con un conocimiento profundo de algunos aspectos de la teoría de la relatividad.

Además, no solamente se acababa de graduar en física en la Facultad de Ciencias Físico-Matemáticas sino que también se había graduado como *Profesor en Ciencias* en la Facultad de Ciencias de la Educación, conducida por el educador Víctor Mercante (1870-1934).[1] Esa Facultad tenía entonces una orientación moderna, de corte cientificista, a la que se agregaba una preocupación seria por los problemas sociales que la Argentina afrontaba en esos años. En su Sección Pedagógica se instaló un laboratorio de psicometría donde, muy tempranamente en la Argentina, se intentó un acercamiento a la nueva tendencia cuantificadora que entonces dominaba en Europa.

Como resultado de la colaboración establecida entre aquellos dos físicos apareció el *Tratado Elemental de Física*, cuyo primer volumen fue publicado en 1928 (Loyarte y Loedel Palumbo, 1928) por la casa editora de Ángel Estrada.[2] Esta editorial, fundada en 1870, era una de las empresas más sólidamente establecidas en el Buenos Aires de esos años. Un segundo volumen, que completaba la obra, apareció cuatro años más tarde (Loyarte y Loedel Palumbo, 1932).

El *Tratado* de Loyarte y Loedel Palumbo alcanzó un éxito editorial considerable y continuó publicándose hasta por lo menos 1964, cuando había alcanzado ya la décima edición. Entre la aparición de ambos volúmenes hubo cambios importantes en las vidas y en las actividades de sus dos autores, y también en sus relaciones personales (von Reichenbach y Andrini, 2015).

A fines de 1927 Loyarte fue electo presidente de la UNLP; con esa elección la nueva universidad pasaba a ser regida, por vez primera, por uno de sus antiguos alumnos. Para Loyarte esa designación abriría un camino que lo identificaría, aún más estrechamente que en el pasado, con la actividad política a nivel nacional, como representante parlamentario del Partido Conservador. Sin duda, la actividad política restó tiempo para sus actividades científicas.

Entre las medidas positivas auspiciadas por Loyarte como presidente de la UNLP se destaca la creación, por primera vez en su universidad, de un número muy reducido de becas externas a las que podían optar los graduados universitarios más destacados. Esta era una antigua ambición de Loyarte, de la que hay evidencias desde antes de su elección como presidente de la Universidad.

---

[1] En 1906 se agregó una *Sección Pedagógica* a la *Facultad de Derecho*, que se transformó en la *Facultad de Ciencias de la Educación* en 1914. Más tarde, en 1920 pasó a denominarse *Facultad de Humanidades y Ciencias de la Educación*. La *Sección Pedagógica* inició la publicación de la revista *Archivos de Pedagogía y Ciencias Afines,* denominada *Archivos de Ciencias de la Educación* a partir de 1915.
[2] El análisis detallado y una lista exhaustiva de textos y manuales de física de nivel secundario utilizados en la Argentina de esos años puede hallarse en Cornejo y López Arriazu (2009).

Loedel Palumbo recibió en 1928 una de esas contadas becas, lo que le permitió trasladarse a Alemania para ampliar su formación científica (Gangui y Ortiz, 2020). Sin duda, en su elección para esa beca, Loedel Palumbo debe haber contado con la aprobación de Loyarte en su doble carácter de presidente de la Universidad y director del Instituto de Física.

El segundo volumen del *Tratado Elemental de Física* se publicó luego de que Loyarte terminara su mandato presidencial y de que Loedel Palumbo regresara de Alemania. Ello sugiere que este volumen fue redactado, y corregido, por lo menos en parte, en un período de intensa actividad política de Loyarte y, también, de conflictos en el campo de sus investigaciones científicas.

Posteriormente ambos autores, aunque ahora en forma separada, continuaron preocupándose por la enseñanza de la física. Sin embargo, lo hicieron en áreas diferentes: Loedel Palumbo profundizó su trabajo sobre los textos de la enseñanza media y, más tarde, se ocupó también del difícil problema de ayudar a los profesores secundarios de esa disciplina a mejorar sus métodos de enseñanza. Loyarte, en cambio, dio preferencia a los textos de enseñanza universitaria de la física, aunque sin dejar de colaborar con otros autores en la redacción de textos de matemáticas para la educación secundaria.

Los intereses pedagógicos de Loyarte y de Loedel Palumbo a partir de la década de 1920 no representan un caso aislado dentro de la comunidad de científicos formada en el Instituto de Física de La Plata (Ortiz y Rubinstein, 2009). Otros miembros del pequeño grupo de físicos graduados entre 1915 y 1925, contribuyeron también a la redacción de textos modernos, adaptados a los programas de diferentes centros educativos, en un momento de expansión de la enseñanza media.

Además, tanto Loyarte como sus colegas José B. Collo (1897-1968) y Teófilo Isnardi (1890-1966) expresaron un interés muy claro por la pedagogía, que se manifestó a través de su inscripción en el profesorado en ciencias o en colaboraciones en *Archivos de Pedagogía*. Sin duda, esa era también una medida de seguridad frente a un porvenir profesional incierto.

También Isnardi y Collo fueron autores de exitosos textos de física (y también de matemática) inicialmente redactados para uso de sus alumnos en la Escuela Naval Militar (Isnardi y Collo, 1938). Esa misma institución se encargó de editarlos a partir de 1924; allí contribuyeron, decididamente, a modernizar la enseñanza de las ciencias exactas (Puglisi, 2011). Más tarde, varios de los volúmenes de esta obra de Isnardi y Collo, en particular los dedicados a la Óptica y a la Electricidad, fueron también usados como material de referencia en los cursos de física que esos autores dictaban en la Facultad de Ciencias de la Universidad de Buenos Aires, especialmente por los estudiantes de ciencias químicas. También fueron usados como texto de física en otras instituciones de la Argentina, o de otros países de América del Sur.

**CONTRIBUCIONES DE LOEDEL PALUMBO A LA ENSEÑANZA DE LA FÍSICA: SUS LIBROS DE TEXTO**

El análisis de la enseñanza de la física a un nivel elemental y medio, es decir, el proceso de transmisión de los resultados de esa ciencia a la juventud, fue una de las preocupaciones principales de Loedel Palumbo a lo largo de su extensa carrera docente. Su temprano acercamiento a la Facultad de Humanidades donde, como hemos dicho, hizo estudios formales de pedagogía y didáctica simultáneamente con sus estudios de física en la Facultad de Ciencias, evidencia esas preocupaciones, aun en su época de estudiante.

Como potencial autor de libros de texto de física para la juventud ese complejo entrenamiento lo colocó en una posición de privilegio. Sin embargo, el esfuerzo considerable que dedicó más adelante a la redacción de obras de física elemental, y también de didáctica de la física al nivel de la enseñanza secundaria, no puede atribuirse exclusivamente a sus intereses intelectuales. Había, sin duda, otras consideraciones de cierto peso, que tienen que ver, tanto con la dinámica de las relaciones personales entre los miembros del Instituto de Física de La Plata, como con la frágil situación de las posiciones en la Universidad frente a la inestable política argentina de esos años (Gangui y Ortiz, 2020).

Sin embargo, independientemente de las razones que motivaron a esos científicos a emprender esas tareas, sus obras de texto dejaron una marca clara en el proceso de actualización de la enseñanza de la física en la Argentina, tanto por su originalidad y por su coherencia lógica, como por su persistente búsqueda de una fundamentación racional.

Hacia fines de la década de 1930 Loedel Palumbo comenzó a dedicar una atención preferente a la redacción de textos para la enseñanza secundaria. Como hemos visto, se había iniciado ya en este género de obras con el exitoso Tratado de física para la escuela media que había escrito en colaboración con Loyarte, y que terminó de publicarse en 1932.

En la primera mitad de la década de 1940, Loedel Palumbo publicó unas 15 obras de texto para la enseñanza secundaria de la matemática, de la física y de la astronomía, a las que imprimió un enfoque muy personal; sus textos tuvieron un éxito editorial considerable.

Un buen número de las obras de Loedel Palumbo llevan el sello de la editorial Ángel Estrada, que continuaba siendo una de las empresas más cotizadas del país en el ramo de los textos de enseñanza. Estrada fue también la editorial que dio un impulso considerable a la difusión de los *Apuntes de Física* de Ricaldoni, una vez que éstos adquirieron una justa reputación.

Más tarde, otra empresa editorial de prestigio, la Editorial Kapelusz, fundada en Buenos Aires en 1905 por el librero austríaco, luego editor, Adolfo Kapelusz (1873-1947), se encargó de publicar otras obras de texto de Loedel Palumbo. Durante muchos años esta última editorial fue la competidora tradicional de Estrada en el mercado argentino de libros de texto.

En la década de 1940 ambas editoriales experimentaron un desarrollo considerable a causa de la dificultad de importar textos de Europa. Ellas lograron penetrar profundamente en el mercado de libros de texto de América Latina y aún de España, particularmente en los años inmediatamente posteriores al fin de la Guerra Civil. Según (Rivera, 1982), las editoriales argentinas llegaron a proveer hasta un 80% de los libros de texto usados en España en esos años.

En esa misma década Loedel Palumbo escribió, en total, más de cuatro mil páginas de obras de texto, donde desarrolló en forma práctica algunas de sus ideas sobre la enseñanza de las ciencias exactas. Entre las obras de texto publicadas en esos años se destaca su *Física elemental* (Loedel Palumbo, 1940a), editada por Estrada, que fue recibida con satisfacción por la crítica local, señalándose en una de las reseñas de la gran prensa diaria (*La Prensa*, 1941) que el autor no había escatimado esfuerzo "en su propósito didáctico". Además, esa nota destacaba otros méritos: había mostrado el "lugar que corresponde al conocimiento de la física en el acervo cultural". Sin duda, esa fue una preocupación central en las obras de Loedel Palumbo.

En esos años la Editorial Kapelusz adquirió recursos técnicos muy modernos que le permitían hacer un uso amplio de ilustraciones, diagramas y fotografías, tanto en negro como en color, en las obras de su fondo editorial. Loedel Palumbo no desaprovechó esas facilidades en la presentación de los libros de texto que publicó en Kapelusz. Además, en esos textos trató de ofrecer una perspectiva más amplia que la que requerían los programas oficiales, agregando notas históricas y perspectivas de las ideas más modernas de la física. En la Argentina ese enfoque no estaba presente en la mayor parte de las obras de texto de esa época.

Nuestro original y prolífico autor también continuó su relación con Estrada: en esos mismos años esa editorial publicó una original obra de Loedel Palumbo, que se titula *Masa y Peso: comedia didáctica en tres actos* (Loedel Palumbo, 1940b). Como indica su título, en esa obra se intenta introducir esos dos conceptos en forma dialogada; es una obra única en su género en la Argentina de esos años.

También en 1940, Estrada publicó *Cosmografía o Elementos de Astronomía* (Loedel Palumbo y de Luca, 1940) en colaboración con Salvador de Luca, profesor de Cosmografía en el Colegio Nacional de La Plata, donde Loedel Palumbo enseñaba física. Este es un texto de más de 600 páginas que tuvo un éxito considerable.

La primera edición de *Cosmografía* fue lanzada en 1940 y, a lo largo de esa década, fue reeditada casi de año en año. A partir de 1941 los mismos autores publicaron, principalmente en Estrada, una serie de textos de matemática elemental que, finalmente, cubrieron los diferentes capítulos de la geometría, la aritmética y el álgebra de la escuela secundaria. La extensa lista de las publicaciones educativas de Loedel Palumbo emula la de Ricaldoni, aunque sin sobrepasarlo.

## ENTRE LA FÍSICA Y LA FILOSOFÍA: UN ANÁLISIS CRÍTICO DE LA ENSEÑANZA DE LA FÍSICA

El interés firme de Loedel Palumbo por la filosofía de las ciencias lo llevó a vincularse con el *Grupo Argentino de Historia de la Ciencia*, una institución local ligada a la *Académie Internationale d'Histoire des Sciences*, cuya sede estaba en París. Ese *Grupo* había sido creado en Buenos Aires en 1933 por iniciativa de Julio Rey Pastor (1888-1962) y de Aldo Mieli (1879-1950) (Ortiz y Pyenson, 1984, 82-85). Los nuevos miembros de ese *Grupo* eran cuidadosamente seleccionados y elegidos por el voto mayoritario de aquellos ya incorporados.

Mieli era un destacado historiador de las ciencias de origen italiano que, antes de buscar refugio en la Argentina, a causa de la Segunda Guerra Mundial, había contribuido a crear en París la antes citada *Académie Internationale*. En Argentina Mieli enseñó historia de las ciencias en un nuevo *Instituto de Historia y Filosofía de la Ciencia,* creado por iniciativa de Rey Pastor en la rama de Santa Fe de la Universidad Nacional del Litoral. Lamentablemente, ese Instituto fue clausurado por las autoridades universitarias impuestas en la Universidad luego del golpe militar de junio de 1943. Quizás el más destacado de los historiadores de la ciencia entrenados en ese instituto fue José Babini (1897-1984), un discípulo de Rey Pastor, que también perdió su cátedra en esa misma ocasión.

En 1935, poco después de la creación del *Grupo Argentino*, Loedel Palumbo se trasladó a París donde participó en el *Congrès International de Philosophie Scientifique*. Años más tarde, en 1949, publicó un extenso e interesante estudio crítico sobre la enseñanza de su disciplina: *Enseñanza de la física* (Loedel Palumbo, 1949), una obra original que entrelaza la física y la pedagogía y sugiere contactos con la historia y la filosofía de la ciencia (Vacaro, 2010).

En su texto, a lo largo de 15 capítulos, Loedel Palumbo se ocupó sucesivamente del método en la enseñanza, la experimentación, la interconexión entre diferentes conceptos científicos y el significado de *teoría física*. Estudió diversos conceptos centrales de la física, como los de masa y temperatura, y analizó, en cierto detalle, la noción de causalidad. Finalmente consideró el papel que esta disciplina debería ocupar en la enseñanza, y la necesidad de incluir en ella referencias a las teorías modernas de la física. Además, intentó caracterizar las diferentes modalidades de los estudiantes de física que había encontrado en sus años de profesor.

En su obra, Loedel Palumbo consideró también el papel auxiliar que puede jugar la historia de la física en la enseñanza de esa ciencia y se ocupó del uso de recursos didácticos al alcance del profesor: láminas, dibujos, diagramas, modelos mecánicos y eléctricos, etc. Su libro es un intento de abrir escenarios nuevos, tanto para los profesores como para los alumnos. El hecho de que una obra de esa naturaleza haya sido escrita, y publicada, en la Argentina de fines de la década de 1940 sugiere que en esos años existían ya núcleos de intelectuales con una preocupación seria y activa por la renovación de la didáctica de las ciencias.

Sin embargo, aunque el interés de Loedel Palumbo por los fundamentos, la historia, la didáctica y la filosofía de las ciencias fue compartido con otros científicos, sería difícil señalar un antecedente específico de *Enseñanza de la física* en la literatura científico-didáctica de la Argentina de la primera mitad del siglo XX. Quizás *El Joven Coleccionista de Historia Natural* de Holmberg, (Holmberg, 1915), escrito 30 años antes para los jóvenes naturalistas, sea una de las pocas obras argentinas que abordan, con cierto éxito, ese enfoque.

En *Enseñanza de la física* Loedel Palumbo no eludió la polémica con sus colegas contemporáneos. Por ejemplo, cuando consideró el carácter de las magnitudes físicas y la temperatura, destacó que algunos de sus colegas locales, por ejemplo, Rey Pastor, Isnardi y Collo, sostenían puntos de vista diversos de los suyos; luego analizó, con cierto detalle, las ideas de Rey Pastor. En trabajos anteriores Loedel Palumbo se había ocupado ya de ese tema, que revisitó en 1948 en la comunicación que leyó ante el Primer Congreso Nacional de Filosofía, celebrado en Mendoza, Argentina.

**CONTRIBUCIONES A LA ENSEÑANZA DE LA FÍSICA A NIVEL UNIVERSITARIO BÁSICO: LA *FÍSICA GENERAL* DE LOYARTE**

Si bien Loyarte contribuyó a la enseñanza de la física a nivel secundario, su aporte en esa área no fue tan extenso y relevante como el de Loedel Palumbo. Sin embargo, fue autor de una *Aritmética* para primero y segundo año y de una *Aritmética y álgebra* para el tercer año del Ciclo Común a los estudios del Bachillerato y del Magisterio. Además, escribió una *Aritmética y Cálculo Mercantil* para el primer año y textos de *Geometría* para el primero y segundo años de la Escuela Superior de Comercio. Todos sus textos fueron escritos en colaboración con el matemático platense Alberto E. Sagastume Berra (1905-1960) y editados por Estrada. Hacia fines de los años 1920 Sagastume había colaborado con otro egresado del Instituto de Física, Rafael Grinfeld (1902-1969), en la redacción de un curso sobre mecánica cuántica dictado por Loyarte en La Plata (Sagastume Berra y Grinfeld, 1928).

Sin embargo, la obra de texto con la que Loyarte dejó una marca muy particular fue su *Física General*, adaptada al curso del mismo nombre que él dictaba en la UNLP. *Física General* comenzó a publicarse en 1922 y fue seguida de otros tres volúmenes. La obra completa se publicó a lo largo de una década y, a partir de entonces, diferentes volúmenes fueron reimpresos varias veces (Loyarte, 1922-1935).

El primer volumen de *Física General* fue reeditado en 1927, y en él Loyarte se ocupó de la *Mecánica del cuerpo rígido; Gravitación; Estática de la elasticidad de los sólidos*. En el segundo volumen se ocupó de la *Estática de los fluidos; Hidrodinámica; Dinámica de la Elasticidad;*[3] *Acústica*. En el tercero consideró el *Calor* (editado por lo menos en 1938), y en el cuarto y último volumen el foco estuvo en *Electricidad y magnetismo; Atomismo de la electricidad;*[4] *Radioactividad; Estructura del átomo; Transmutación artificial de los elementos*. Este último volumen tuvo una edición en 1935-36, una quinta en 1944 y, más tarde, alcanzó por lo menos a una décima edición.

Con esa obra Loyarte cubrió un arco extenso, que abarca desde la mecánica clásica hasta los resultados más recientes de la física nuclear. Presentó las diferentes materias en una forma clara y accesible; ocasionalmente hizo uso de diagramas y fotografías que contribuían a facilitar la comprensión de los temas tratados. No faltan ya en esa obra referencias a trabajos de investigación publicados recientemente en revistas internacionales.

---

[3] En algunas ediciones se agrega al título: "*Mecánica de los medios continuos*".
[4] En algunas ediciones se agrega al título: "*Electroquímica*".

Puede decirse que *Física General* tuvo un impacto considerable en la enseñanza universitaria de la física, ya que permitió a los alumnos de La Plata llegar, en su propio idioma, hasta los umbrales mismos de los descubrimientos contemporáneos.

## RECEPCIÓN CONTEMPORÁNEA DE *FÍSICA GENERAL*

Debido a su amplia aceptación en los diferentes centros universitarios, la influencia de *Física General* se hizo sentir, no solamente en La Plata, sino también en Buenos Aires y en varias otras ciudades de Latino América en cuyas universidades fue, durante largos años, el texto oficial o una respetable obra de referencia.

Las reseñas que se publicaron sobre esa obra sugieren que fue percibida como un libro de lectura agradable, con excelentes diagramas y descripciones de experimentos interesantes. Un comentarista local del primer volumen, el Ing. Manuel Ucha (Ucha, 1922), colega de Loyarte, indica que, como ya lo anunciara el autor en el prólogo, *Física General* supone en sus lectores cierta familiaridad con los principios del cálculo diferencial e integral y de las ecuaciones diferenciales con coeficientes constantes, temas que el autor usa en varias partes de su obra.

En la referencia de esa misma obra publicada en *Revista de Matemática y Física Elementales*,[5] dirigida por el ingeniero y matemático Bernardo Baidaff (1889?-1967), se destaca que se trata de "la única fuente donde los estudiantes universitarios [argentinos] podrán hallar, en lengua materna, una exposición metódica, clara y precisa de la física general" (Baidaff, 1922, 18).

Si bien para mediados de la década de 1930 esa obra había alcanzado ya su cuarto y último volumen, varias referencias de la época, y aun posteriores, indican erróneamente que la obra constaba de cinco volúmenes. Es posible que Loyarte haya considerado la posibilidad de agregar un quinto volumen, posiblemente sobre *Óptica*, tema que en parte quedó fuera de su obra; en la nota antes citada Baidaff alude a esa posibilidad.

En 1945, luego del fallecimiento de Loyarte, Loedel Palumbo comenzó la publicación de un tratado moderno de óptica. La Facultad de Ciencias Fisicomatemáticas de La Plata imprimió sus notas, que luego fueron reeditadas por el Centro de Estudiantes de Ingeniería de La Plata. Es posible que este tratado haya sido pensado como una conclusión natural de la *Física General* de Loyarte. Sin embargo, la conflictiva situación política de ese período (von

---

[5] Loyarte había enviado interesantes colaboraciones a esa revista.

Reichenbach y Bibiloni, 2012), y luego la separación de Loedel Palumbo como Profesor de la Universidad de La Plata, impidieron la continuación de esta publicación, de la que sólo alcanzó a imprimirse el primer fascículo (Loedel Palumbo, 1945).

*Física General* fue también referenciada en revistas extranjeras; la nota de la revista inglesa *The Mathematical Gazette* (1934) destaca tanto la calidad de sus diagramas e ilustraciones como la de sus descripciones de experimentos. Sin embargo, lamenta la carencia de ejercicios y de ejemplos que permitan al alumno evaluar su comprensión de la teoría, lo que es exacto. Finalmente, en el marco internacional de textos de física general, el autor de la nota destaca que la estructura de ese libro no es diferente de la de otros tratados de uso internacional, y lo califica como "an uninspired textbook".

Debemos recordar que en esos años existían ya en el mercado internacional varias obras considerablemente más avanzadas y modernas que *Física General*: por ejemplo, el *Cours de physique générale* de Georges Bruhat (1887-1945) (Bruhat, 1924-34), publicado en aproximadamente los mismos años que la obra de Loyarte, aunque no en el mismo orden.[6] En 1936 Bruhat le agregó un volumen específicamente de problemas: *Recueil de problèmes*.

## CONTRIBUCIONES A LA ENSEÑANZA DE LA FÍSICA A NIVEL UNIVERSITARIO AVANZADO: LA *FÍSICA RELATIVISTA* DE LOEDEL PALUMBO

Antes de concluir nuestro trabajo, volvamos a Loedel Palumbo, cuya *Física relativista* (Loedel Palumbo, 1955) se publicó en Buenos Aires en 1955. Esta es una obra extensa, de nivel universitario más avanzado que la obra de Loyarte, difundida por la Editorial Kapelusz en una cuidada edición.

En esa obra Loedel Palumbo volvió a ocuparse de la teoría de la relatividad, tema al que años antes había contribuido con diversos trabajos de investigación (Gangui y Ortiz, 2020). Según explica en la introducción, esa obra fue su contribución personal a la celebración del cincuentenario de la aparición del primer trabajo de Einstein sobre la teoría de la relatividad restringida.

*Física relativista* es una obra accesible a un público universitario con conocimientos básicos de matemáticas y de física, en la que el autor desarrolla un estudio histórico-crítico de la teoría de la relatividad. Además, presenta algunas novedades; una de ellas es una original representación gráfica de ciertas transformaciones entre sistemas de referencia en movimiento relativo, hoy conocidas como *Diagramas de Loedel Palumbo.*

---

[6] El volumen *Électricité* apareció en 1924; *Thermodynamique* en 1926; *Optique* en 1930 y *Mécanique* en 1934.

En su libro comienza considerando las relaciones entre la física clásica y la física relativista y el impacto de la nueva teoría sobre la concepción tradicional del espacio y del tiempo. Se ocupa luego de la cinemática y la dinámica relativistas, y también de las pruebas experimentales de la teoría de la relatividad; esta última parte ocupa aproximadamente un tercio de la obra.

A continuación, el texto de Loedel Palumbo considera la posibilidad de extender la teoría restringida de la relatividad, para lo cual, en el capítulo siguiente, introduce a sus lectores en los elementos del cálculo tensorial. Con esas herramientas estudia las ecuaciones del campo gravitatorio, las pruebas experimentales de la teoría de la gravitación de Einstein y el principio de la velocidad parabólica, concluyendo con un capítulo sobre Cosmología.

La obra de Loedel Palumbo fue recibida favorablemente por los críticos locales; uno de ellos, Babini, a quién hemos aludido más atrás, expresó que se trataba de "un excelente y útil tratado para el conocimiento y comprensión" de una parte importante de las conquistas de la física del siglo XX (Babini, 1956). Otra nota bibliográfica, también muy favorable, apareció sin firma en el diario *La Nación* (*La Nación*, 1956). Sin embargo, otros críticos consideraron que se trataba de una presentación muy personal de la teoría: es posible que éste haya sido, precisamente, el principal interés de esa obra.

**IMPACTO INTERNACIONAL DE *FÍSICA RELATIVISTA* DE LOEDEL PALUMBO**

El texto de relatividad de Loedel Palumbo fue, quizás, recibido con una mayor atención en el extranjero que en la Argentina. Mientras que otros trabajos del autor, y su debate con Loyarte acerca de la rotación cuantificada del átomo de mercurio (von Reichenbach y Andrini, 2015) se circunscribieron principalmente a la Argentina, con un impacto muy limitado fuera del país, el trabajo de Loedel Palumbo sobre aspectos geométricos de la teoría de la relatividad tuvo un impacto internacional amplio.

En 1964 el físico teórico Hans-Jürgen Treder (1928-2006), de la Academia de Ciencias de Berlín, publicó una extensa e inteligente reseña del libro de Loedel Palumbo sobre la teoría de la relatividad en el *Zentralblatt für Mathematik und ihre Grenzgebiete* (Treder, 1964). Señaló Treder que el punto central de ese libro no era, solamente, cubrir el desarrollo matemático de la teoría de la relatividad, sino que, además, su autor trataba de ofrecer una presentación detallada de sus fundamentos físicos.

Indicó también que la postura del autor frente a la teoría de la relatividad tenía afinidades con la de Georges Birkhoff (1884-1944), que había propuesto formulaciones alternativas de dicha teoría. Es muy posible que Loedel Palumbo haya asistido a las conferencias que Birkhoff dictó en Buenos Aires, en La Plata, en Córdoba, y en otras ciudades argentinas durante una visita extensa que realizó a ese país en 1942 (Ortiz, 2003). En algunas de esas conferencias, por ejemplo, en las de Córdoba, se ocupó específicamente de su formulación de la relatividad.

Agudamente, Treder indicó también que el libro de Loedel Palumbo contenía un capítulo que "va más allá del mero contenido de un texto y que está basado en su trabajo original" (Treder, 1964, 217). Se refería a su propuesta alternativa a los célebres diagramas de Minkowski en el marco de la relatividad especial que, como hemos señalado, fueron llamados los *Diagramas* de Loedel Palumbo.

**CONCLUSIONES**

A partir de 1925, las contribuciones pedagógicas de los físicos de la UNLP muestran un abanico amplio de intereses que va desde las obras de carácter elemental, para la escuela secundaria, hasta las que atienden a las necesidades de los estudiantes universitarios que cursan estudios básicos en ramas de las ciencias exactas y de la ingeniería.

Además, algunos de esos científicos expresaron un interés activo por la historia y la filosofía de la física, y se preocuparon por contribuir al estudio histórico-crítico del proceso de transmisión de la física en la enseñanza media.

En este trabajo hemos delineado brevemente el largo camino que separa y enlaza los textos escritos por varios físicos de la Argentina en la primera mitad del siglo XX. Una primera etapa, que en la Argentina podemos identificar como la *era de Ricaldoni*, se caracteriza por un enfoque puramente fenomenológico de la física, con un marcado énfasis en la descripción de fenómenos físicos, experimentos e instrumentos, sin olvidar notas capaces de captar la imaginación de sus lectores con un espíritu que recuerda a Jules Verne (von Reichenbach y Cappannini, 2016). Esa práctica, que había sido corriente en Europa unas décadas antes, dominó en la Argentina desde fines del siglo XIX hasta fines de la Primera Guerra Mundial. La obra paradigmática fue los *Apuntes de Física* de Ricaldoni, en sus múltiples versiones desde 1896 hasta los años 1920.

Sin duda, en la Argentina la transición hacia contenidos más modernos no fue completa y estuvo condicionada por los dictados de los programas oficiales de física, que decidían sobre el posible éxito de comercialización de esas obras. La serie de obras de texto posterior al modelo de Ricaldoni, que en alguna forma había sido inspirado por la obra clásica de Adolphe Ganot (1804-1887), (Ganot, 1851), pretendió atribuir un énfasis más marcado a los principios fundamentales de las ciencias físicas.

Interesa destacar que los autores de esa segunda serie de obras de texto fueron, en forma dominante, estudiosos locales entrenados por los físicos alemanes Emil Bose y Richard Gans en el Instituto de Física de La Plata, dentro del primer cuarto del siglo XX.

La construcción, instalación y mantenimiento del Instituto de Física a lo largo de la primera mitad del siglo XX -inicialmente bajo la dirección de Ricaldoni, luego bajo la dirección de los dos físicos alemanes arriba nombrados y más tarde de Loyarte- implicó una contribución económica substancial por parte del Estado Nacional. Argumentamos en este trabajo que la aportación a la cultura científica, a nivel nacional, de algunos de los textos de física mencionados más atrás se destaca como uno de los logros más significativos, y socialmente más claramente tangibles, del esfuerzo económico volcado por el Estado argentino en la instalación y mantenimiento del Instituto de Física de La Plata.

## BIBLIOGRAFÍA

Babini, José, (1956), 'Un gran tratado sobre "Física Relativista" de Enrique Loedel'. *El Mundo*, Buenos Aires, julio 8, 1956.

Baidaff, Bernardo (1922), 'Bibliografía', *Revista de Matemática y Física Elementales*, 4, pp. 18-19.

Bruhat, Georges (1924-1934), *Cours de physique générale* : *à l'usage de l'enseignement supérieur scientifique et technique.* Paris: Masson et Cie.

Cámara de Diputados (1926), *Diario de Sesiones de la Honorable Cámara de Diputados de la Nación*, 1926, No. 5.

Cornejo, Jorge; López Arriazu, Francisco (2009), "La enseñanza de la Física en la Escuela Media Argentina (1880-1930): un análisis desde los manuales escolares". *Revista Electrónica de Enseñanza de las Ciencias*, 9 (1), pp. 326-341.

Gangui, Alejandro; Ortiz, Eduardo L. (2008), "Einstein's Unpublished Opening lecture for his Course on Relativity Theory in Argentina, 1925", *Science in Context*, 21 (3), pp. 435-450.

Gangui, Alejandro; Ortiz, Eduardo L. (2014), "The scientific impact of Einstein's visit to Argentina, in 1925", *Asian Journal of Physics*, 23 (1-2), pp. 81-90. [arxiv.org/abs/1603.03792]

Gangui, Alejandro; Ortiz, Eduardo L. (2020), "El físico Enrique Loedel Palumbo en el corredor científico Montevideo-Buenos Aires-La Plata: 1920-1930", *Revista de Indias*, 80 (280), pp. 815-846.

Ganot, Adolphe (1851), *Traité élémentaire de physique expérimentale et appliquée et de météorologie*. Paris: Chez L'Auteur. Traducido al Español en 1860, con el título de: *Tratado elemental de física experimental y aplicada, y de meteorología, con una selecta colección de problemas, adornado con 586 bellos grabados de madera intercalados en el texto.* París: Librería de Rosa y Bouret; Madrid: Librería de Aug. Bouret.

Holmberg, Eduardo L. (1915), *El joven coleccionista de historia natural en la República Argentina*, Buenos Aires: Sociedad Luz.

Isnardi, Teófilo; Collo, José B. (1938), *Curso de Física*. Río Santiago: Escuela Naval Militar. (En cuatro tomos editados a partir de 1924.)

La Nación (1956), 'Física relativista', 06.05.1956.

La Prensa (1941), 'Textos Escolares', 09.04.1941.

Loedel Palumbo, Enrique (1940a), *Física elemental*. Buenos Aires: Ángel Estrada.

Loedel Palumbo, Enrique (1940b), *Masa y Peso: comedia didáctica en tres actos*. Buenos Aires: Estrada.

Loedel Palumbo, Enrique (1945), *Tratado de Óptica. Fascículo 1*. La Plata: Facultad de Ciencias Fisicomatemáticas. (Reeditado por el Centro de Estudiantes de Ingeniería, La Plata).

Loedel Palumbo, Enrique (1949), *Enseñanza de la física*. Buenos Aires: Kapelusz (Volumen 4 de la 'Biblioteca de Ciencias de la Educación', 520 pp.).

Loedel Palumbo, Enrique (1955), *Física Relativista*. Buenos Aires: Kapelusz.

Loedel Palumbo, Enrique; de Luca, Salvador, (1940), *Cosmografía o Elementos de Astronomía*. Buenos Aires: Ángel Estrada.

Loyarte, Ramón G. (1922-1935), *Física General*, I-IV. La Plata: UNLP.

Loyarte, Ramón G.; Loedel Palumbo, Enrique (1928), *Tratado elemental de física. Tomo I (responde ampliamente a los programas de los colegios nacionales y escuelas normales),* Buenos Aires: Ángel Estrada Editores, primera edición.

Loyarte, Ramón G.; Loedel Palumbo, Enrique (1932), *Tratado elemental de física. Tomo II (responde ampliamente a los programas de los colegios nacionales y escuelas normales),* Buenos Aires: Ángel Estrada Editores, primera edición.

Nazar Anchorena, Benito A. (1927), *La Universidad Nacional de La Plata en el año 1926 (Presidencia del Dr. Benito A. Nazar Anchorena),* Buenos Aires: Peuser.

Ortiz, Eduardo L. (1995), "A convergence of Interests: Einstein's visit to Argentina en 1925". *Ibero-Amerikanisches Archives*, 21 (1-2), pp. 67-126.

Ortiz, Eduardo L. (2003), "La política interamericana de Roosevelt: George D. Birkhoff y la inclusión de América Latina en las redes matemáticas internacionales". *Saber y tiempo*, 4 (15), pp. 53-111 y 4 (16), pp. 21-70.

Ortiz, Eduardo L.; Pyenson, Lewis (1984), "José Babini, matemático e historiador de las ciencias", *Llull*, 7, pp. 77-98.

Ortiz, Eduardo L.; Rubinstein, Héctor (2009), "La Física en la Argentina en los dos primeros tercios del siglo veinte: algunos condicionantes exteriores a su desarrollo". *Revista Brasileira de História da Ciência*, 2 (1), pp. 40-81.
https://www.sbhc.org.br/arquivo/download?ID_ARQUIVO=43

Puglisi, Alfio A. (2011), "La enseñanza de la física en la Escuela Naval Militar", *Revista de Publicaciones Navales*, Tomo CXXXX, Año CXII, Nro. 707, primer cuatrimestre, pp. 29-43.
https://xdoc.mx/documents/la-enseanza-de-la-fisica-en-la-escuela-naval-5dd2fb5a5448c

Puiggrós, Adriana (2002), *Qué pasó en la educación argentina*, Buenos Aires: Galerna, cap. II.

Pyenson, Lewis (1985), *Cultural Imperialism and Exact Sciences, German Expansion Overseas 1900-1930*. New York: Peter Lang.

Ricaldoni, Tebaldo J. (1896), *Apuntes de Física*. Buenos Aires: Instituto Vertíz.

Rivera, Jorge B. (1982), *El auge de la industria cultural (1930-1955),* En AA.VV, *Historia de la literatura argentina*, (2ª ed.), 4, pp. 577-600.

Saavedra Lamas, C. (1916), *Reformas Orgánicas en la República Argentina. Sus antecedentes y fundamentos*. Tomo I y II. Buenos Aires: Imprenta Argentina Jacobo Peuser.

Sagastume Berra, Alberto E.; Grinfeld, Rafael (1928), "Mecánica atómica". *Anales de la Sociedad Científica Argentina*, 105, 1928, pp. 11-42; 106, 1928, pp. 7-24, pp. 159-177; 109, 1930, pp. 209-238.

Solari, Manuel H. (2002), *Historia De La Educación Argentina*. Buenos Aires: Paidós.

The Mathematical Gazette (1934), "Bibliography", 18 (231), p. 360 (Dic. 1934).

Treder, Hans-Jürgen (1964), "Loedel, Enrique: Relativistische Physik". *Zentralblatt für Mathematik und ihre Grenzgebiete*, 68, pp. 216-217.
https://www.digizeitschriften.de/en/dms/toc/?PID=PPN245319514_0068

Ucha, Manuel (1922), "Física General", Tomo I, por R. G. Loyarte, *ASCA*, 93, pp. 284-85.

Vacaro, Daniel (2010), Resumen: Los hilos de la trama conceptual. Enrique Loedel Palumbo y la enseñanza de la Física. *V Jornadas de Historia de la Ciencia Argentina (2009)*, Buenos Aires: UNTREF.

von Reichenbach, María Cecilia; Andrini, Leandro (2015), "Una nueva forma de energía cuantificada: presentación de la polémica Loyarte Loedel". *Saber y Tiempo electrónica* (Buenos Aires), 1, pp. 169-185.

von Reichenbach, María Cecilia; Bibiloni, Aníbal Guillermo (2012), "Las dificultades de implantar una disciplina científica. Los primeros cincuenta años del Instituto de Física de La Plata". En: Hurtado de Mendoza, Diego (ed.), *La Física y los físicos argentinos, historias para el presente*, Córdoba: Universidad Nacional de Córdoba.

von Reichenbach, María Cecilia; Cappannini, Osvaldo (2016), "Similitudes y diferencias entre dos propuestas dominantes de la enseñanza de las ciencias en la Argentina en las primeras décadas del siglo XX: Adolphe Ganot y Tebaldo Ricaldoni", *Revista Brasileira de História da Ciência*, 9 (1), pp. 6-18.

Zusman, Perla (1996), "Una geografía científica para ser enseñada. La Sociedad Argentina de Estudios Geográficos (1922-1940)", *Doc. Anal. Geogr.*, 31, pp. 171-189.